\documentclass{scrartcl}

\usepackage{bm}
\usepackage{url}
\usepackage{amsmath}
\usepackage{amssymb}
\usepackage{graphicx}

\title{Theoretical study of reactions in the three body  $e^-e^+\bar{p}$ system and  antihydrogen formation cross sections}

\author{V.\,A.\,Gradusov\/\thanks{e-mail: v.gradusov@spbu.ru}, V.\,A.\,Roudnev\/\thanks{e-mail: v.rudnev@spbu.ru}, E.\,A.\,Yarevsky\/\thanks{e-mail: e.yarevsky@spbu.ru}, S.\,L.\,Yakovlev\/\thanks{e-mail: s.yakovlev@spbu.ru}
}

\date{
Department of Computational Physics, St Petersburg State University, St Petersburg 199034, Russia\\
\today
}

\begin{document}

\maketitle

\abstract{
We apply a new highly efficient method of solving Faddeev-Merkuriev equations to multi-channel scattering calculations of the antihydrogen formation cross section for antiproton scattering off the ground and excited states of the positronium. Our results demonstrate good agreement with the known data on total and partial cross sections for all the reaction channels. Using moderate computational resources we have achieved a supreme energy resolution. 
}

Several experiments on antimatter based on the use of the Antiproton Decelerator facility are being planned and performed at CERN.
Two of them aimed at the antimatter gravitational behaviour --- AEgIS~\cite{AEGIS15} and GBAR~\cite{GBAR15} --- use, inter alia, the three-body reaction
\begin{equation}
\label{react}
\bar{p}+\mathrm{Ps}\to\overline{\mathrm{H}}+e^-
\end{equation}
of antihydrogen $\overline{\mathrm{H}}$ formation via antiproton $\bar{p}$ scattering off the gas of Rydberg positronium (Ps) to obtain the antimatter species. Even though charge exchange reactions have a long history of experimental and theoretical studying, only a few approaches demonstrated some degree of success in studying the $e^-e^+\bar{p}$ scattering in multi-channel deep inelastic regime~\cite{Hu02,Hu02-2,Lazau18, Kadyr17, Krasn19}. In particular, the authors of~\cite{Kadyr17, Krasn19} have discussed the growth of the antihydrogen formation cross sections occuring just above the highly excited Ps thresholds, which is of special interest as a mechanism of enhancement of the antihydrogen formation reaction rate when producing antimatter atoms.

Theoretical and computational investigation of the reaction, however, is complicated by the presence of multiple near-threshold resonances, long-range polarization interactions in multiple channels, complex contributions of multiple virtual excitations of various geometries. As an example of the delicate nature of the system we could mention the Gailtis-Damburg oscillations~\cite{Gaili63, Gaili63-2} that originate from the long-range dipole interaction between the excited neutral atom (either $\overline{\mathrm{H}}$ or Ps) and the charged particle ($e^-$ or $\bar{p}$). All these features of the system make dimensionality reduction problematic and call for some approach, which would take into account the fully dimensional dynamics of the system.

Because of this complex nature of the system, the number of reliable theoretical results remains limited: the known results lack either energy or cross section resolution, or both. There is a definitive lack of reliable data on the low-energy $e^-e^+\bar{p}$ scattering cross sections.

We have proposed and implemented an approach for solving the quantum three-body problem which combines both a theoretically sound technique and a computationally efficient algorithm~\cite{Grad21}.
It is based on solving the Faddeev-Merkuriev (FM) equations~\cite{Fadd93} in total orbital momentum representation~\cite{Kostr89}.
Our earlier calculations~\cite{Grad21} have shown that our approach makes for highly precise calculations of the bound state energies for the states with high total orbital momentum.
Here we apply it to scattering problems and perform a series of calculations of antihydrogen formation cross sections for the reaction~(\ref{react}).
We also compare our results with the published antihydrogen formation cross sections~\cite{Hu02,Hu02-2,Lazau18} for the antiproton scattering off the ground as well as the first excited states of Ps.

We consider the system of three spinless nonrelativistic charged particles of masses $m_{\alpha}$ and charges $Z_\alpha$, $\alpha=1,2,3$. In what follows the set of indices $\{\alpha$, $\beta$, $\gamma\}$ runs over the set $\{1,2,3\}$ enumerating particles.
By pair $\alpha$ we call a pair of particles $\beta\gamma$ complementary to particle $\alpha$.
Particle positions are described by the set of coordinates.
In the center of mass frame, the standard choice is the set of Jacobi coordinates.
They are defined for a partition $\alpha(\beta\gamma)$ as relative position vectors between the particles of the pair $\alpha$, and between their center of mass and the particle $\alpha$.
It is convenient to use reduced Jacobi coordinates $(\bm{x}_\alpha, \bm{y}_\alpha)$ which are Jacobi vectors scaled by factors $\sqrt{2\mu_{\alpha}}$ and $\sqrt{2\mu_{\alpha(\beta\gamma)}}$, respectively.
Here the reduced masses are given by
\begin{equation}
\mu_{\alpha} = \frac{m_\beta m_\gamma}{m_\beta+m_\gamma},\qquad \mu_{\alpha(\beta\gamma)}=\frac{m_\alpha (m_\beta+m_\gamma)}{m_\alpha+m_\beta+m_\gamma}.
\end{equation}
For different $\alpha's$ the reduced Jacobi vectors are related by an orthogonal transform $\bm{x}_\beta=c_{\beta\alpha}\bm{x}_\alpha + s_{\beta\alpha}\bm{y}_\alpha$, $\bm{y}_\beta=-s_{\beta\alpha}\bm{x}_\alpha + c_{\beta\alpha}\bm{y}_\alpha$~\cite{Fadd93}.
In what follows where it is due, it is assumed that $\beta$ Jacobi vectors are represented through $\alpha$.

In the reduced Jacobi coordinates the FM equations for three charged particles~\cite{Fadd93, merkur80} can be written as (the bold font is used for vectors)
\begin{multline}
\label{MFeq}
\{ T_\alpha  +  V_\alpha(x_\alpha) +
V_\beta^{(\mathrm{l})}(x_\beta,y_\beta)+V_3(x_3,y_3) - E \} \psi_\alpha(\bm{x}_\alpha, \bm{y}_\alpha) \\
  = - V_\alpha^{(\mathrm{s})}(x_\alpha,y_\alpha) \psi_\beta(\bm{x}_\beta, \bm{y}_\beta),\quad
\alpha\ne\beta=1,2.
\end{multline}
Here $T_\alpha\equiv-\Delta_{\bm{x_\alpha}} - \Delta_{\bm{y_\alpha}}$ are the kinetic energy operators.
The potentials $V_\alpha$ represent the pairwise Coulomb interaction $V_{\alpha}(x_\alpha)=\sqrt{2\mu_{\alpha}}Z_\beta Z_\gamma/x_\alpha$ ($\beta,\gamma\ne\alpha$).
It is assumed that the potential $V_3$ is repulsive.
The potentials $V_\alpha$ are split into the interior (short-range) $V^{(\mathrm{s})}_\alpha$ and the tail (long-range) parts $V^{(\mathrm{l})}_\alpha$
\begin{equation}
\label{PotSplit}
V_\alpha(x_\alpha) = V^{(\mathrm{s})}_\alpha(x_\alpha,y_\alpha) + V^{(\mathrm{l})}_\alpha(x_\alpha, y_\alpha).
\end{equation}
The equations~(\ref{MFeq}) can be summed up leading to the Schr\"odinger equation for the wave-function
$\Psi=\sum_{\alpha}\psi_\alpha$, where $\psi_{\alpha}$ are the components of the wave function given by the solution of the equations~(\ref{MFeq}).

The splitting~(\ref{PotSplit}) is done according to $V_\alpha^{(\mathrm{s})}(x_\alpha,y_\alpha) = \chi_\alpha(x_\alpha, y_\alpha) V_\alpha(x_\alpha)$, where the Merkuriev cut-off function $\chi_\alpha$ confines the short-range part of the potential to the regions in the three-body configuration space corresponding to the three-body collision point and the binary configuration ($x_\alpha \ll y_\alpha$, when $y_\alpha \to \infty$)~\cite{Fadd93}.
Following~\cite{Grad16,Grad19}, we use the cut-off function in the two-body configuration space of pair $\alpha$
\begin{equation}
\label{Mcutoff}
\chi_\alpha(x_\alpha, y_\alpha) = \chi_\alpha(x_\alpha) = 2/\left\{1+\exp[ (x_\alpha/x_{0\alpha})^{2.01} ]\right\}.
\end{equation}
The parameter $x_{0\alpha}$ can in principle be chosen arbitrarily, but its choice changes the properties of components $\psi_\alpha$ that are important from both the theoretical and computational points of view~\cite{Yak-Papp10}.
It can be effectively chosen by the simple practical algorithm presented in~\cite{Grad19}.

The splitting procedure makes the properties of the FM equations for Coulomb potentials as appropriate for  scattering problems as standard Faddeev equations in the case of short-range potentials \cite{Papp-Yak01}. The key property  of the FM equations~(\ref{MFeq})
is that the right-hand side of each equation is confined to the vicinity of the triple collision point \cite{Yak-Papp10}. It results in the asymptotic uncoupling of the
set of FM equations and, therefore, the asymptote of each component $\psi_\alpha$ for energies below the breakup threshold contains only terms corresponding to binary configurations of pairing $\alpha$  \cite{Yak-Papp10, Papp-Yak01}.
For the total energy $E$ of the system below the three-body ionization threshold the asymptote reads
\begin{equation}
\label{asympt3D}
\psi_\alpha(\bm{x}_\alpha,\bm{y}_\alpha) = \chi_{\mathbb{A}_0}(\bm{x}_\alpha,\bm{y}_\alpha)\delta_{\mathbb{A}\mathbb{A}_0}+\Xi_\alpha(\bm{x}_\alpha,\bm{y}_\alpha),
\end{equation}
where the outgoing wave is of the form
\begin{equation}
\label{xi-def}
\Xi_\alpha(\bm{x}_\alpha,\bm{y}_\alpha) =  \sum\limits_{n\ell m}\frac{\phi_A(x_\alpha)}{x_\alpha}Y_{\ell m}(\hat{\bm{x}}_\alpha)\sqrt{\frac{p_{n_0}}{p_n}} \widetilde{\mathcal{A}}_{\mathbb{A},\mathbb{A}_0}(\hat{\bm{y}}_\alpha, \bm{p}_{n_0})\frac{\mathrm{e}^{\mathrm{i}(p_n y_\alpha-\eta_n\log(2p_n y_\alpha))}}{y_\alpha}.
\end{equation}
Here $Y_{\ell m}$ stands for the standard spherical harmonic function.
The multi-index $\mathbb{A}=\{Am\}=\{\alpha n\ell m\}$ specifies the scattering channels, i.e. various two-body Coulomb bound states in the pair $\alpha$ with the wave function $\phi_{A}(x_\alpha)Y_{\ell m}(\hat{\bm{x}}_\alpha)/x_\alpha$ and the energy $\varepsilon_n$.
The momentum $p_n$ of the outgoing particle
is determined by the energy conservation condition $E=p_n^2+\varepsilon_n$, and the Sommerfeld parameter is defined as  $\eta_n\equiv Z_\alpha(Z_\beta+Z_\gamma)\sqrt{2m_{\alpha(\beta\gamma)}}/(2p_n)$.
The initial channel is specified by the incoming wave
\begin{eqnarray}
\label{chi-def}
\chi_{\mathbb{A}_0}(\bm{x}_\alpha,\bm{y}_\alpha) &=& \frac{\phi_{A_0}(x_\alpha)}{x_\alpha}Y_{\ell_0m_0}(\hat{\bm{x}}_\alpha)
\mathrm{e}^{\mathrm{i}(\bm{p_{n_0}},\bm{y_\alpha})} \mathrm{e}^{-\pi\eta_{n_0}/2} \nonumber \\
& \times & \Gamma(1+\mathrm{i}\eta_{n_0}) {}_1F_1(-\mathrm{i}\eta_{n_0}, 1, \mathrm{i}(p_{n_0}y_\alpha-(\bm{p}_{n_0},\bm{y}_\alpha))),
\end{eqnarray}
where ${}_1F_1$ is the confluent hypergeometric function~\cite{NISTDLMF}.
The binary scattering amplitude
\begin{equation}
\mathcal{A}_{\mathbb{A}\mathbb{A}_0}(\hat{\bm{y}}_\alpha, \bm{p}_{n_0})= \mathcal{A}_{\mathrm{C}}(\hat{\bm{y}}_\alpha, \bm{p}_{n_0}) + \widetilde{\mathcal{A}}_{\mathbb{A},\mathbb{A}_0}(\hat{\bm{y}}_\alpha, \bm{p}_{n_0})
\end{equation}
corresponds to the transition from the initial binary channel $\mathbb{A}_0$ to the binary channel $\mathbb{A}$.
Here $\mathcal{A}_{\mathrm{C}}$ is the standard two-body Coulomb scattering amplitude~\cite{Mess61}.
The scattering cross section is given by
\begin{eqnarray}
\label{cs}
\sigma_{AA_0} = \frac{1}{2m_{\alpha_0(\beta\gamma)}(2\ell_0+1)} \sum_{m=-\ell}^{\ell}\sum_{m_0=-\ell_0}^{\ell_0}\int\text{d}\hat{\bm{y}}_\alpha\left|\mathcal{A}_{\mathbb{A}\mathbb{A}_0}(\hat{\bm{y}}_\alpha, \bm{p}_{A_0})\right|^2.
\end{eqnarray}

By adding the boundary conditions~(\ref{asympt3D})-(\ref{chi-def}) to the FM equations~(\ref{MFeq}), one obtains the boundary-value problem, which can be solved numerically.
However, each equation~(\ref{MFeq}) is a six-dimensional partial differential equation.
To make calculations possible, the equations are reduced by projecting~(\ref{MFeq}) onto a subspace of a given total angular momentum~\cite{Kostr89}, which is an integral of motion for the processes considered here.
To this end, one introduces the more appropriate kinematic coordinates ($X_\alpha$, $\Omega_\alpha$) in the six dimensional configuration space of the problem.
The coordinates $X_\alpha=\{x_\alpha,y_\alpha,z_\alpha\equiv
 (\bm{x_\alpha},\bm{y_\alpha})/(x_\alpha y_\alpha)\}$ determine particle positions in the plane containing them.
The remaining three coordinates $\Omega_\alpha=\{\phi_\alpha,\vartheta_\alpha,\varphi_\alpha\}$ determine the position of the plane in the space.
They are the standard Euler angles of a rotation of some laboratory system of coordinates to the body-fixed system of coordinates~\cite{Varsh89} in which the vector $\bm{x}_\alpha$ is positioned along the z-axis and the vector $\bm{y}_\alpha$ lies in the right half of the xz-plane.
The FM components in the new coordinates are expanded as
\begin{equation}
\label{exp-full}
\psi_\alpha(X_\alpha,\Omega_\alpha) = \sum_{L=0}^{+\infty}\sum_{\tau=\pm1}\sum_{M=-L}^{L}\sum_{M'=M_0}^{L} (1-z_\alpha^2)^{M'/2}
\frac{\psi_{\alpha MM'}^{L\tau}(X_\alpha)}{x_\alpha y_\alpha}
F_{MM'}^{L\tau}(\Omega_\alpha).
\end{equation}
Here $M_0 = (1-\tau)/2$, the functions
\begin{equation}
\label{F-def}
F_{MM'}^{L\tau}(\Omega_\alpha)=
\frac{1}{\sqrt{2+2\delta_{M'0}}} \left( D_{MM'}^L(\Omega_\alpha)+\tau(-1)^{M'}D_{M,-M'}^L(\Omega_\alpha) \right),
\end{equation}
are linear combinations of Wigner $D$-functions $D_{MM'}^L$~\cite{Varsh89,Bieden81}.
The function $F_{MM'}^{L\tau}$ is the common eigenfunction of the total orbital momentum squared, its projection and the spatial inversion operators~\cite{Kostr89,Bieden81} with eigenvalues $L$, $M$ and $\tau$.
The multiplier $(1-z_\alpha^2)^{M'/2}$ in~(\ref{exp-full}) is introduced to make the partial components $\psi_{\alpha MM'}^{L\tau}$ and their derivatives nonsingular at $z_\alpha=\pm1$~\cite{Grad21,Scrinzi96}.
Now substituting the series~(\ref{exp-full}) into the FM equations~(\ref{MFeq}) written in new coordinates ($X_\alpha$, $\Omega_\alpha$) and projecting the resulting equations onto the functions $F_{MM'}^{L\tau}$,
one gets the finite set of 3D equations for partial components~$\psi_{\alpha MM'}^{L\tau}(X_\alpha)$
\begin{multline}
\label{FM}
\big[
T_{\alpha MM'}^{L\tau}+V_{\alpha}(x_\alpha)+
V_{\beta}^{(\mathrm{l})}(x_\beta,y_\beta) 
+V_3(x_3,y_3)-E
\big]
\psi_{\alpha MM'}^{L\tau}(X_\alpha) \\
+
T_{\alpha M,M'-1}^{L\tau-} \psi_{\alpha M,M'-1}^{L\tau}(X_\alpha)+
T_{\alpha M,M'+1}^{L\tau+} \psi_{\alpha M,M'+1}^{L\tau}(X_\alpha) \\
 = -\frac{V_{\alpha}^{(\mathrm{s})}(x_\alpha,y_\alpha)}{(1-z_\alpha^2)^{\frac{M'}{2}}}
\frac{x_\alpha y_\alpha}{x_\beta y_\beta}
\sum_{M''=M_0}^{L} \frac{(-1)^{M''-M'}2}{\sqrt{2+2\delta_{M''0}}} \\
\times F_{M''M'}^{L\tau}(0,w_{\beta\alpha},0)
(1-z_\beta^2)^{\frac{M''}{2}}
\psi_{\beta MM''}^{L\tau}(X_\beta).
\end{multline}
The kinetic part is of the form 
\begin{multline}
T_{\alpha MM'}^{L\tau} = -\frac{\partial^2}{\partial y_\alpha^2} +
\frac{1}{y_\alpha^2}\left( L(L+1) - 2M'^2 \right) - 
\frac{\partial^2}{\partial x_\alpha^2} \\ 
 - \left( \frac{1}{y_\alpha^2}+\frac{1}{x_\alpha^2} \right)
\bigg(
(1-z_\alpha^2)\frac{\partial^2}{\partial z_\alpha^2}-
2(M'+1)z_\alpha\frac{\partial}{\partial z_\alpha}
 -
M'(M'+1)
\bigg),
\end{multline}
\begin{eqnarray}
T_{\alpha M,M'+1}^{L\tau+}  &=&
\frac{1}{y_\alpha^2}
\lambda^{L,M'}\sqrt{1+\delta_{M'0}} \left[
-(1-z_\alpha^2)\frac{\partial}{\partial z_\alpha}+
2(M'+1)z_\alpha
\right], \nonumber \\
T_{\alpha M,M'-1}^{L\tau-}  &=&
\frac{1}{y_\alpha^2}
\lambda^{L,-M'}\sqrt{1+\delta_{M'1}}
\frac{\partial}{\partial z_\alpha}.
\end{eqnarray}
Here $\lambda^{LM'}=\sqrt{L(L+1)-M'(M'+1)}$.
The kinematic angle $w_{\beta\alpha}$ is related to the transform of coordinates $(X_\alpha,\Omega_\alpha)$ with different $\alpha$ and is given by
\begin{equation}
w_{\beta\alpha}=
\left\{
\begin{array}{l}
\arccos\frac{-s_{\beta\alpha}y_\alpha z_\alpha+c_{\beta\alpha}x_\alpha}{x_\beta},\ \mathrm{if}\ (\beta-\alpha)\ \mathrm{mod}\ 3=2,\\
2\pi-\arccos\frac{-s_{\beta\alpha}y_\alpha z_\alpha+c_{\beta\alpha}x_\alpha}{x_\beta},\ \mathrm{otherwise},
\end{array}
\right.
\end{equation}
where the range of $\arccos$ is $[0,\pi]$.
The obtained equations are the (3D) FM equations in total 
orbital momentum representation.
The most important property of the system~(\ref{FM}) is that the equations on partial components $\psi_{\alpha MM'}^{L\tau}$ with different indices $L$, $M$ and $\tau$ form independent sets of equations. This is the direct consequence of the fact that for the three-body systems considered here the total orbital momentum, its projection and the spatial parity are conserved.
For given $L$, $M$ and $\tau$ the system~(\ref{FM}) consists of $3(L-M_0+1)$ three-dimensional PDEs.
The partial components $\psi_{\alpha MM'}^{L\tau}$ must satisfy zero Dirichlet-type boundary conditions on the lines $x_\alpha=0$, $y_\alpha = 0$.

The boundary conditions on the partial components $\psi_{\alpha MM'}^{L\tau}$ takes the form of the sum
\begin{equation}
\label{asympt-part}
\psi_{\alpha MM'}^{L\tau}(X_\alpha)=\chi_{\mathbb{A}_0 MM'}^{L\tau}(X_\alpha)\delta_{\mathbb{A}\mathbb{A}_0}+\Xi_{\alpha MM'}^{L\tau}(X_\alpha),
\end{equation}
of the partial components of the incoming and outgoing waves defined in~(\ref{asympt3D})-(\ref{chi-def}).
They can be obtained by projecting~(\ref{xi-def}) and~(\ref{chi-def}) onto the functions $F_{MM'}^{L\tau}$.
If the laboratory system of coordinates is chosen so that the vector $\bm{p}_{n_0}$ is positioned along its z axis, the incoming wave partial component is given by
\begin{multline}
\label{chi-fin}
\chi_{\mathbb{A}_0 MM'}^{L}(X_\alpha)=\delta_{-M,m_0}\frac{(-1)^M\sqrt{(2\ell_0+1)}}{p_{n_0}\sqrt{2+2\delta_{M'0}}}
\phi_{A_0}(x_\alpha)  \\
\times\sum_{\lambda=|L-\ell_0|}^{L+\ell_0}\sqrt{2\lambda+1}\mathrm{i}^\lambda\mathrm{e}^{\mathrm{i}\sigma_\lambda(\eta_{n_0})}F_\lambda(\eta_{n_0},p_{n_0}y_\alpha) \\
\times\frac{Y_{\lambda M'}(\theta_\alpha,0)}{(1-z_\alpha^2)^{\frac{M'}{2}}}
C_{\lambda,0,\ell_0,M}^{L,M}C_{\lambda,M',\ell_0,0}^{L,M'}\left(1+\tau(-1)^{\lambda+\ell_0-L}\right),
\end{multline}
where the Coulomb phase shift $\sigma_\lambda(\eta_{n_0})=\arg\Gamma(1+\lambda+\mathrm{i}\eta_{n_0})$, $F_\lambda$ is the regular Coulomb function~\cite{Mess61} and $C_{j1,m1,j2,m2}^{j,m}$ are the Clebsch-Gordan coefficients.
The outgoing wave partial component reads
\begin{multline}
\label{xi-fin}
\Xi_{\alpha MM'}^{L\tau}(X_\alpha) = \delta_{M, -m_0}\frac{(-1)^M}{\sqrt{4\pi(2+2\delta_{M'0})}}
 \sum_{n\ell}\sqrt{2\ell+1} \phi_{A}(x_\alpha) 
\mathrm{e}^{\mathrm{i}(p_n y_\alpha-\eta_n\log(2p_n y_\alpha))} \\
\times
\sum_{\lambda=|L-\ell|}^{L+\ell}\frac{Y_{\lambda M'}(\theta_\alpha,0)}{(1-z_\alpha^2)^{\frac{M'}{2}}} \sqrt{\frac{p_{n_0}}{p_n}}\widetilde{\mathcal{A}}_{A\mathbb{A}_0}^{L\lambda}C_{\lambda,M',\ell,0}^{L,M'}\left(1+\tau(-1)^{\lambda+\ell-L}\right).
\end{multline}
The partial amplitude $\widetilde{\mathcal{A}}_{A\mathbb{A}_0}^{L\lambda}$ is related to the coefficients of the expansion of the amplitude $\widetilde{\mathcal{A}}_{\mathbb{A}\mathbb{A}_0}(\hat{\bm{y}}_\alpha)$ in terms of spherical harmonics.
It can be shown that the cross section $\sigma_{AA_0}$ given by~(\ref{cs}) can be expressed by
\begin{eqnarray}
\label{cs-j}
\sigma_{AA_0} &=& \sum_{L=0}^{+\infty}\sigma_{AA_0}^L,\nonumber \\
\sigma_{AA_0}^L &=& \frac{1}{2m_{\alpha_0(\beta\gamma)}}\frac{1}{2\ell_0+1}\sum_{m_0=-\ell_0}^{\ell_0}\sum_{\lambda=|L-\ell|}^{L+\ell}\left|\mathcal{A}_{A\mathbb{A}_0}^{L\lambda}\right|^2,
\end{eqnarray}
where $\sigma_{AA_0}^L$ are the partial cross sections, through the partial total amplitude given by
\begin{equation}
\mathcal{A}_{A\mathbb{A}_0}^{L\lambda} =  \widetilde{\mathcal{A}}_{A\mathbb{A}_0}^{L\lambda}+\delta_{AA_0} \frac{\sqrt{\pi(2\lambda+1)}}{\mathrm{i}p_{n_0}}
\left(\mathrm{e}^{\mathrm{i}2\sigma_\lambda(\eta_{n_0})}-1\right)C_{\lambda,0,\ell_0,-m_0}^{L,-m_0}.
\end{equation}

Subtracting the incoming wave from the FM components, one obtains the driven equations~(\ref{FM}) with the inhomogeneous term.
Its solution must satisfy zero Dirichlet-type boundary conditions on the lines $x_\alpha=0$, $y_\alpha = 0$ and be asymptotically equal to the outgoing wave~(\ref{xi-fin}).
The obtained boundary problem is solved numerically by the spline collocation method.
The scheme is described in~\cite{Grad21} where the interested reader can find the details.
For implementing the outgoing boundary conditions we use a hybrid basis, which is obtained by explicitly adding the outgoing waves to the spline basis set in variable $y_\alpha$.
These additional basis functions have the form of the irregular Coulomb function~\cite{Mess61} $u_{\ell}^+(\eta_{n},p_{n}y_\alpha)$ in the asymptotic region and are polynomials chosen to satisfy the zero Dirichlet-type boundary condition at the origin and to ensure the required continuity of the basis function in the solution interval.
The use of the hybrid basis set on the one hand ensures the fulfillment of outgoing boundary conditions.
On the other hand it reduces the required number of basis functions since the additional functions describe the behaviour of the solution at large $y_\alpha$ quite well.

In the presented results we hold the accuracy of calculations of cross sections within the
error range not exceeding 1\%.
Binary scattering processes are specified by initial and final atom states. For example, ${\mathrm{Ps}(1)\to\overline{\mathrm{H}}(2)}$ denotes an excited $n=2$ (both $s$ and $p$ states) antihydrogen formation process when antiproton is scattering off the ground $n=1$ state of Ps.
In the Ore gap energy region, where direct and rearrangement processes involving ground state antihydrogen and Ps atoms are possible, we have calculated partial cross sections with $L$=0--9.
The partial and summed cross sections are presented in Table~\ref{Ore-tab} and Figure~\ref{Ore-fig} and compared with the results of other authors.
\begin{table}
\centering
\begin{tabular}{c|c|c|c|c}
\hline
$E$, a.u. &-0.22947 & -0.21832 & -0.17955 & -0.13828 \\
\hline
$\sigma^{L\le4}_{\mathrm{Ps}(1)\to\mathrm{Ps}(1)}$ & 22.1 & 20.6 & 18.8 & 17.4 \\
\cite{Hu99} & 21.95 & 20.64 & 18.88 & 17.23 \\
\cite{Gien97} & 22.00 & 20.57 & 19.16 & 18.20 \\
\hline
$\sigma^{L\le4}_{\mathrm{Ps}(1)\to\overline{\mathrm{H}}(1)}$ &  & 3.31
 & 3.81 & 4.02 \\
\cite{Hu01} &  & 3.2943 & 3.7858 & 4.0551 \\
\cite{Gien97} &  & 3.250 & 3.779 & 4.076 \\
\hline
$\sigma^{L\le9}_{\mathrm{Ps}(1)\to\overline{\mathrm{H}}(1)}$ &  & 3.31 & 3.82 & 4.07 \\
\cite{Hu01} &  & 3.2949 & 3.9795 & 4.1043 \\
\hline
\end{tabular}
\caption{Total cross sections (in units of $\pi a_0^2$) summed up to the specified maximum value of the total momentum $L$ in the Ore gap compared with the results of other authors.}
\label{Ore-tab}
\end{table}
\begin{figure}
\includegraphics[width=1.0\textwidth]{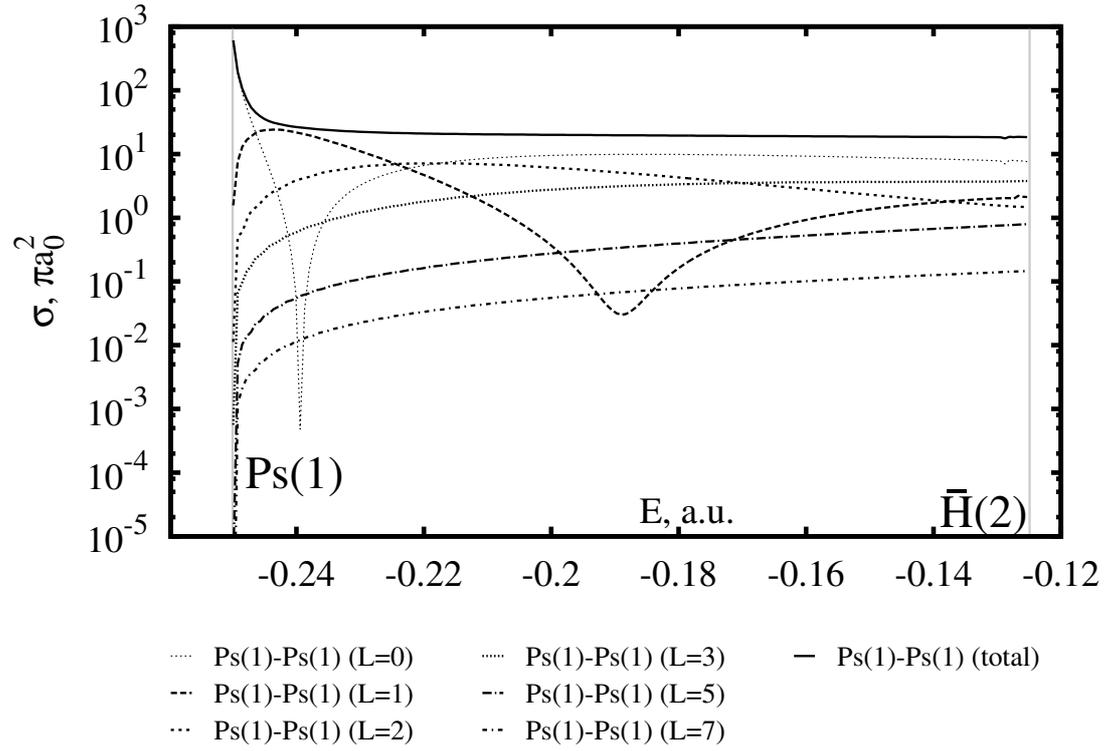}
\caption{Partial and total elastic cross sections (in units of $\pi a_0^2$) in Ps channel in the Ore gap. The total cross section is obtained by summing up the partial cross sections with $L$=0--9.}
\label{Ore-fig}
\end{figure}
\begin{figure}
\includegraphics[width=1.0\textwidth]{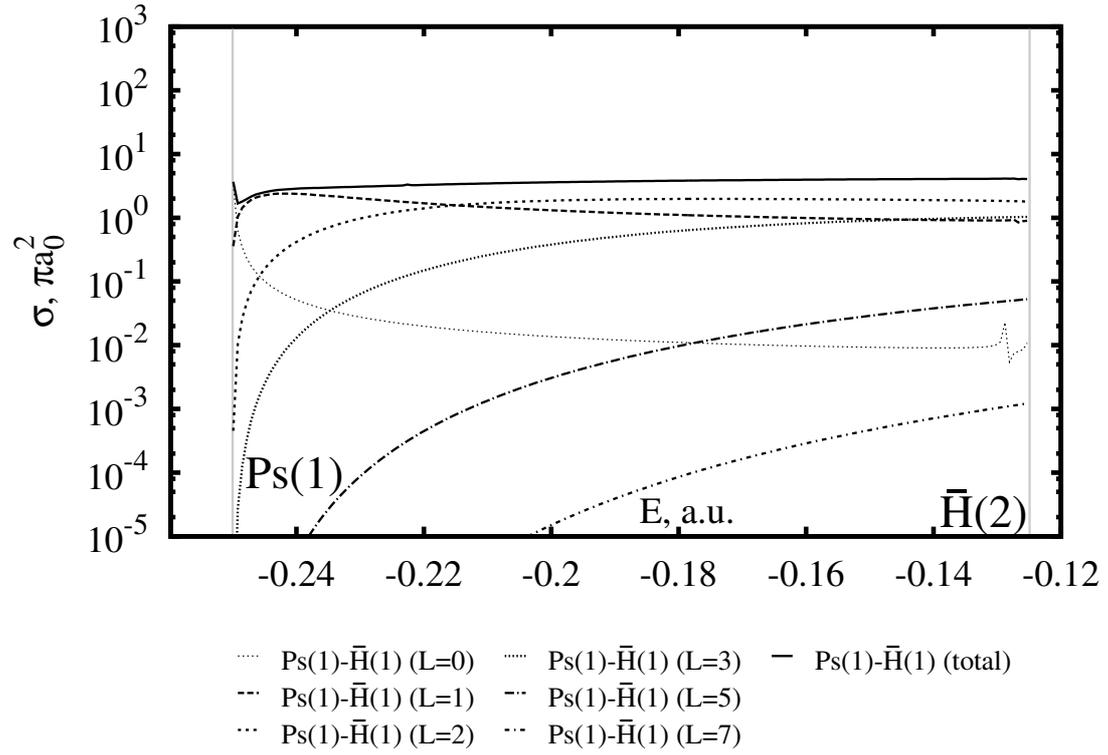}
\caption{Partial and total rearrangement cross sections (in units of $\pi a_0^2$) of antihydrogen formation in the Ore gap. The total cross section is obtained by summing up the partial cross sections with $L$=0--9.}
\label{Ore-fig}
\end{figure}

Comparison of our results with results of other authors in the energy region above the excited antihydrogen $\overline{\mathrm{H}}(2)$ threshold is given in Tables~\ref{H2-tab} and~\ref{Ps2-tab}.
\begin{table}
\centering
\begin{tabular}{c|c|c|c|c}
\hline
$E$, a.u. & -0.11473 & -0.09973  & -0.08473 & -0.07973 \\
\hline
$\sigma^0_{\mathrm{Ps}(1)\to\mathrm{Ps}(1)}$ & 7.10 & 6.44 & 5.82 & 5.63 \\
\cite{Hu99} & 7.09 & 6.44 & 5.83 & 5.63 \\
\cite{Mitroy95} &  & 6.45 &  &  \\
\hline
$\sigma^1_{\mathrm{Ps}(1)\to\mathrm{Ps}(1)}$ & 2.26 & 2.53 & 2.79 & 2.87 \\
\cite{Hu99} & 2.28 & 2.54 & 2.64 & 2.87 \\
\cite{Mitroy95} &  & 2.51 &  &  \\
\hline
$\sigma^2_{\mathrm{Ps}(1)\to\mathrm{Ps}(1)}$ & 1.24 & 1.03 & 0.862 & 0.817 \\
\cite{Hu99} & 1.16 & 1.01 & 0.929 & 0.790 \\
\cite{Mitroy95} &  & 1.02 &  &  \\
\hline
$\sigma^0_{\mathrm{Ps}(1)\to\overline{\mathrm{H}}(1)}$ & 0.00801 & 0.00758 & 0.00719 & 0.00704 \\
\cite{Hu99} & 0.00815 & 0.00780 & 0.00729 & 0.00715 \\
\hline
$\sigma^1_{\mathrm{Ps}(1)\to\overline{\mathrm{H}}(1)}$ & 0.860 & 0.807 & 0.757 & 0.741 \\
\cite{Hu99} & 0.858 & 0.805 & 0.742 & 0.739 \\
\hline
$\sigma^2_{\mathrm{Ps}(1)\to\overline{\mathrm{H}}(1)}$ & 1.76 & 1.67 & 1.59 & 1.56 \\
\cite{Hu99} & 1.77 & 1.69 & 1.57 & 1.58 \\
\hline
$\sigma^0_{\mathrm{Ps}(1)\to\overline{\mathrm{H}}(2)}$ & 0.0844 & 0.0952 & 0.107 & 0.113 \\
\cite{Hu99} & 0.0884 & 0.0927 & 0.105 & 0.114 \\
\hline
$\sigma^1_{\mathrm{Ps}(1)\to\overline{\mathrm{H}}(2)}$ & 0.273 & 0.630 & 0.854 & 0.908 \\
\cite{Hu99} & 0.268 & 0.632 & 1.05 & 0.910 \\
\hline
\end{tabular}
\caption{Partial cross sections (in units of $\pi a_0^2$) in the $\overline{\mathrm{H}}(2)$--Ps(2) energy region compared with the results of other authors.}
\label{H2-tab}
\end{table}
\begin{table}
\centering
\begin{tabular}{c|c|c|c|c}
\hline
$E$, a.u. & -0.06228 & -0.06198 & -0.06123 & -0.05978 \\
\hline
$\sigma^0_{\mathrm{Ps}(1,2)\to\overline{\mathrm{H}}(1)}$ & 0.169 & 0.078 & 0.037 & 0.022 \\
\cite{Hu02} & 0.282 & 0.097 & 0.047 & 0.030 \\
\hline
$\sigma^1_{\mathrm{Ps}(1,2)\to\overline{\mathrm{H}}(1)}$ & 3.67 & 1.98 & 1.20 & 0.944 \\
\cite{Hu02} & 3.373 & 1.783 & 1.130 & 0.886 \\
\hline
$\sigma^0_{\mathrm{Ps}(1)\to\overline{\mathrm{H}}(2)}$ & 0.106 & 0.105 & 0.104 & 0.103 \\
\cite{Hu02} & 0.125 & 0.116 & 0.112 & 0.107 \\
\hline
$\sigma^1_{\mathrm{Ps}(1)\to\overline{\mathrm{H}}(2)}$ & 0.999 & 0.995 & 0.993 & 0.992 \\
\cite{Hu02} & 1.041 & 1.042 & 1.015 & 1.040 \\
\hline
$\sigma^0_{\mathrm{Ps}(2)\to\overline{\mathrm{H}}(2)}$ & 184 & 79.9 & 34.0 & 16.9 \\
\cite{Hu01} & 218.84 & 76.701 & 32.481 & 17.201 \\
\hline
$\sigma^1_{\mathrm{Ps}(2)\to\overline{\mathrm{H}}(2)}$ & 479 & 229 & 102 & 50.1 \\
\cite{Hu01} & 482.65 & 226.62 & 101.91 & 50.73 \\
\hline
\end{tabular}
\caption{Partial cross sections (in units of $\pi a_0^2$) in the Ps(2)-$\overline{\mathrm{H}}(3)$ energy region compared with the results of other authors.}
\label{Ps2-tab}
\end{table}
Despite the good overall agreement, there are some significant discrepancies in the values of cross sections associated with the excited Ps, especially at energies just above the Ps(2) threshold.
We conclude thus that obtaining accurate excited Ps cross sections is quite a challenging task from both the theoretical and computational points of view.
The difficulties are associated with extended Ps-antiproton interaction region enlarged by both the extended excited Ps wavefunction and long-range dipole interaction between the Ps and antiproton~\cite{Hu99}.

\begin{figure}
\includegraphics[width=1.0\textwidth]{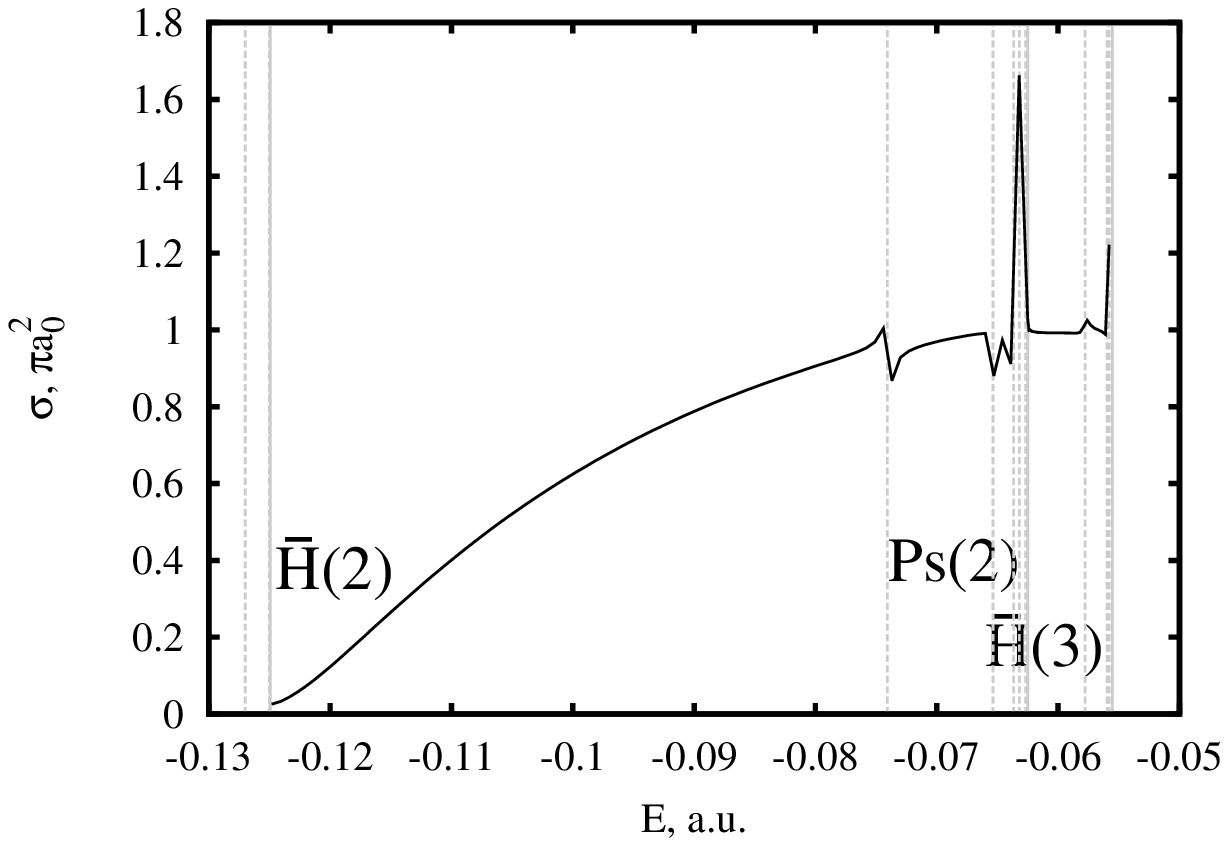}
\caption{Antihydrogen formation cross section $\sigma^1_{\mathrm{Ps}(1)\to\overline{\mathrm{H}}(2)}$. Vertical dashed lines mark positions of resonances~\cite{ho04,varga08,yu12,umair14}.}
\label{a-excit-1}
\end{figure}

\begin{figure}
\includegraphics[width=1.0\textwidth]{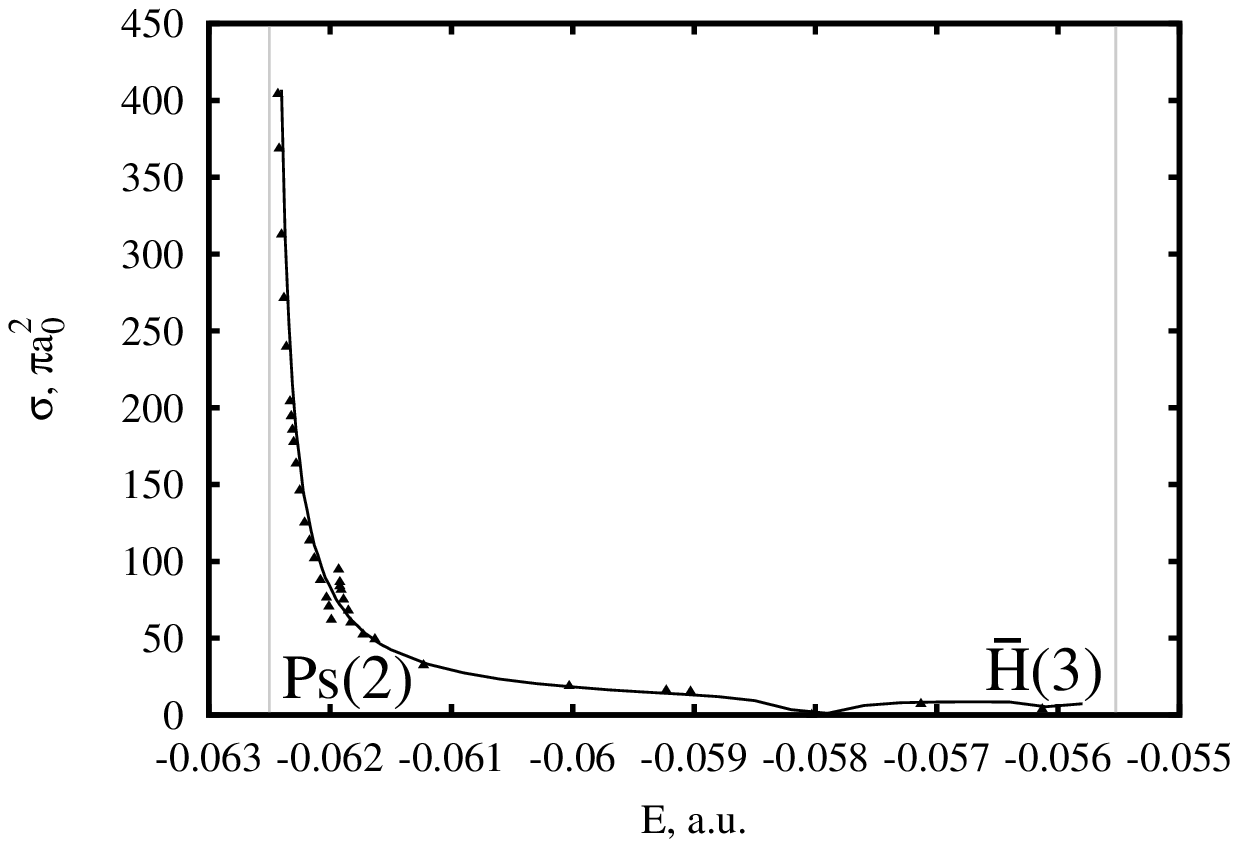}
\caption{Antihydrogen formation cross section $\sigma^0_{\mathrm{Ps}(2)\to\overline{\mathrm{H}}(1,2)}$. Black triangles mark points given in~\cite{Lazau18}.}
\label{a-form-0}
\end{figure}

In Figures~\ref{a-excit-1} and~\ref{a-form-0} we present some antihydrogen formation partial cross sections.
We identify a number of Feshbach resonances in the $\sigma^1_{\mathrm{Ps}(1)\to\overline{\mathrm{H}}(2)}$ cross section.

To summarize, we have calculated the cross sections of antihydrogen formation via the reaction~(\ref{react}) in the energy region above the first excited state of the Ps threshold.
In the future, we are planning to extend our calculations to energy regions where the higher excitations of Ps are possible.  

The research of V.A.G. was supported by Russian Science Foundation grant No. 19-72-00076.
Research was carried out using computational resources provided by Resource Center ``Computer Center of SPbU'' (http://cc.spbu.ru).

\bibliography{refs}

\end{document}